\documentclass[%
 reprint,
 amsmath,amssymb,
 aps,
]{revtex4-2}

\usepackage{graphicx}
\usepackage{dcolumn}
\usepackage{bm}
\usepackage{hyperref}
\hypersetup{colorlinks=true, citecolor=blue, urlcolor=blue, linkcolor=blue}
\usepackage{float}
\usepackage{color}
\usepackage[section]{placeins}

\graphicspath{{figures/}}

\begin{document}

\preprint{APS/123-QED}

\title{Impact of Committed Minorities: Unveiling Critical Mass of Cooperation in the Iterated Prisoner's Dilemma Game}

\author{Zhixue He$^1$}

\author{Chen Shen$^2$}
\email{steven\_shen91@hotmail.com}

\author{Lei Shi$^{1,3}$}
\email{shi\_lei65@hotmail.com}

\author{Jun Tanimoto$^{2,4}$}

\affiliation{
1. School of Statistics and Mathematics, Yunnan University of Finance and Economics, 650221, Kunming, China.\\
2. Faculty of Engineering Sciences, Kyushu University, Fukuoka, 816-8580, Japan\\
3. Interdisciplinary Research Institute of data science, Shanghai Lixin University of Accounting and Finance, 201209, Shanghai, China\\
4. Interdisciplinary Graduate School of Engineering Sciences, Kyushu University, Fukuoka, 816-8580, Japan
}

\date{\today}

\begin{abstract}

The critical mass effect is a prevailing topic in the study of complex systems. Recent research has shown that a minority of zealots can effectively drive widespread cooperation in social dilemma games. However, achieving a critical mass of cooperation in the prisoner's dilemma requires stricter conditions. The underlying mechanism behind this effect remains unclear, particularly in the context of repeated interactions. This paper aims to investigate the influence of a committed minority on cooperation in the Iterated Prisoner's Dilemma game, a widely studied model of repeated interactions between individuals facing a social dilemma. In contrast to previous findings, we identify tipping points for both well-mixed and structured populations. Our findings demonstrate that a committed minority of unconditional cooperators can induce full cooperation under weak imitation conditions. Conversely, a committed minority of conditional cooperators, who often employ Tit-for-Tat or extortion strategies, can promote widespread cooperation under strong imitation conditions. These results hold true across various network topologies and imitation rules, suggesting that critical mass effects may be a universal principle in social dilemma games. Additionally, we discover that an excessive density of committed conditional cooperators can hinder cooperation in structured populations. This research advances our understanding of the role of committed minorities in shaping social behavior and provides valuable insights into cooperation dynamics.
\\ 

\end{abstract}

\maketitle


\section{\label{sec:1}Introduction}

The study of the \textit{critical mass effect} explores significant changes in state and properties that occur within complex nonlinear systems\cite{eckmann1985ergodic,fruchart2021non,achlioptas2009explosive,dorogovtsev2008critical,baxter2015critical}. In natural systems, even minor perturbations near tipping points can trigger can lead to sudden and drastic transformations, often referred to as ``explosive changes''. For instance, the intricate interplay between system structure and dynamic characteristics can result in explosive synchronization\cite{gomez2011explosive}. Similarly, establishing a specific threshold for quarantine probability has been observed to effectively mitigate the spread of epidemics by isolating infected individuals\cite{lagorio2011quarantine}, while a minority of dissenting particles can disrupt the flocking state in active Brownian motion\cite{yllanes2017many}. Remarkably, these critical phenomena also appear in social systems like voting and social segregation. Whether it involves the convergence of social opinion in voter models\cite{mobilia2003does,crokidakis2014impact} or the emergence of distinct segregation patterns characterized by similar preferences, such as observed in Schelling's segregation model\cite{jensen2018giant}, these social phenomena are initiated by a few individuals and propagate from individual behavior to collective behavior. Such critical phenomena provide valuable insights into system dynamics, particularly emphasizing the crucial role of a minority of individuals in driving overall evolution of social systems.

Studying the evolution of human behavior in social dilemmas provides profound insights into the development of social systems, as it involves the inherent conflict between individual interests and collective interests\cite{hofbauer2003evolutionary,szabo2007evolutionary,axelrod1987evolution,axelrod1981evolution,press2012iterated,szabo2007evolutionary}. Pairwise social dilemma games, such as the harmony game, stag-hunt game, snowdrift game, and Prisoner's Dilemma game, capture various forms of conflicting interests and exhibit different equilibrium properties within these game types\cite{szabo2007evolutionary,hofbauer2003evolutionary,szabo2007evolutionary}. In the realm of social dilemma games, the influence of committed individuals, known as ``zealots'', on the emergence of cooperation has been explored using evolutionary game theory \cite{mobilia2007role,mobilia2003does,arendt2015opinions,masuda2012evolution,matsuzawa2016spatial,cardillo2020critical,nakajima2015evolutionary,shen2022simple,sharma2023small,guo2023facilitating}. Research indicates that committed cooperators, even as a minority, can effectively trigger widespread cooperation in stag-hunt dilemma games, which represent coordination problems \cite{cardillo2020critical}. Stag-hunt games have two pure strategy Nash equilibria, and the initial fraction of cooperators in the population heavily influences the outcomes, enabling zealots to more easily induce large-scale cooperation\cite{cardillo2020critical}. Conversely, in snowdrift games, which represent situations where individuals benefit from being different from others (anti-coordination problems), the presence of committed defectors as a minority can facilitate large-scale cooperation in well-mixed populations, but not in structured populations~\cite{guo2023facilitating}. As for Prisoner's Dilemma games, which capture the fundamental conflict between individual and collective interests and are frequently used to address the cooperation conundrum, the emergence of large-scale cooperation depends on factors such as population structure, imitation strength, or update rules, rather than relying solely on the presence of committed cooperators \cite{cardillo2020critical,matsuzawa2016spatial,sharma2023small}. 

However, Previous research has primarily focused on analyzing one-shot game scenarios, where players do not have repeated encounters with the same opponents~\cite{masuda2012evolution,matsuzawa2016spatial,cardillo2020critical,nakajima2015evolutionary,shen2022simple,sharma2023small,guo2023facilitating}. In contrast, in realistic scenarios with repeated interactions, the dynamic nature of the process allows for the emergence of a broader range of strategies that can be adopted by players over time\cite{axelrod1987evolution,axelrod1981evolution,press2012iterated}. One well-known strategy in iterated games is Tit-for-Tat (TFT), where individuals start by cooperating and then mirror their opponent's previous actions in subsequent interactions. Despite its simplicity, TFT has proven highly successful in sustaining cooperation within populations in Iterated Prisoner's Dilemma (IPD) games\cite{axelrod1987evolution,axelrod1981evolution}. Expanding on the concept of iterated games, Press and Dyson introduced a class of strategies known as zero-determined (ZD) strategies\cite{press2012iterated}, which establish a linear relationship between an individual's payoff and their opponent's payoff. Extortion strategy, a subset of ZD strategies, enable players to achieve higher payoffs than their opponents by a certain percentage\cite{hilbe2013evolution}, and can serve as a catalyst for promoting cooperation under specific conditions or in certain population scenarios\cite{adami2013evolutionary,hilbe2014extortion,xu2017extortion,wu2014boosting,szolnoki2014evolution}. 

Building upon this, the question arises: Can zealots initiate a universal critical mass effect in the Prisoner's Dilemma game regardless of factors like select intensity, network topologies, or update rules when repeated interactions are considered? To investigate this, we concentrate on the Prisoner's Dilemma game, which poses a significant challenge for the emergence of cooperation, instead of examining the critical mass of cooperation in games such as stag-hunt and snowdrift. Our objective is to understand the mechanisms behind how a small group of committed individuals can trigger widespread cooperation within the context of the IPD game. We focus on three typical strategies in IPD game: \textit{unconditional cooperation}, \textit{unconditional defection} and \textit{extortion}, as well as taking into account the presence of committed individuals who adopt one of these strategies. Our findings reveal that even a small number of committed individuals can effectively trigger widespread collective cooperation in both well-mixed and structured populations, surpassing the scope of a one-shot Prisoner's Dilemma game. Interestingly, we have discovered that committed extortioners, under conditions of strong selection, exhibit a significant critical mass effect that enhances cooperation. Similarly, committed cooperators demonstrate a similar critical mass effect in a weak selection scenario. Additionally, we evaluate the robustness of this phenomenon by analyzing different network topologies and alternative strategy updating rules, thereby confirming the impact of committed individuals on the critical mass effect of cooperation. In conclusion, these results imply that the critical mass of cooperation could serve as a universal principle in realistic social dilemma games, as it remains resilient against network topologies, network structures, imitation strength, and update rules.

\section{\label{sec:2} Model}

\paragraph*{\textbf{Game model and strategies setting}}
Donation game is a paradigm of IPD game, in which a player interacts with the same opponent multiple times. In each interaction, both players either cooperate or defect. A cooperator pays a cost $c$ to benefit her opponent with $b$ ($b>c>0$), while a defector does nothing, resulting in the player receiving reward $R$ (punishment, $P$) from mutual cooperation (mutual defection) and obtaining the temptation $T$ (suckers' payoff, $S$) from defection (cooperation) when unilateral defection occurs. In an iterated game, the one-memory strategy is considered, which is represented by a four-element tuple $(p_R,p_S,p_T,p_P)$ that describes the probability an individual chooses to cooperate based on the outcome of the previous interaction. Some typical strategies can be written as: \textit{unconditional cooperation} [$C$ = (1,1,1,1)], \textit{unconditional defection} [$D$ = (0,0,0,0)], and Tit-for-Tat [TFT = (1,0,1,0)]. Then, let $p_R = p_T =:p$ and $p_P = p_S =:q$, through an analysis of various combinations of $p$ and $q$ and their resulting strategy interactions, Ref.\cite{hilbe2013evolution} established the concept of $\chi$-extortion strategy $E\chi$, where extortion factor $\chi$ means that the extortioner's surplus exceeds the opponent's surplus with  $\chi$-fold ($\chi>1$). The TFT strategy has been demonstrated to be one of the most successful strategies in a wide range of scenarios\cite{axelrod1981evolution,axelrod1987evolution,hilbe2014extortion}, leading to its widespread utilization and discussion. Notably, the TFT strategy is equivalent to the extortion strategy when $\chi=1$\cite{hilbe2013evolution}. Therefore, we primarily focus on the three strategies: $C$, $D$, and $E\chi$ strategy. From the perspective of the average payoff of long-term interaction, the payoff matrix can be written as: 
\begin{equation} 
\label{eq:pm}
\centering
  \bordermatrix{
    & C & D &  E_\chi \cr
  C & b-c & -c &  \frac{b^2-c^2}{b\chi +c}  \cr
  D & b & 0 & 0  \cr
 E_\chi & \frac {(b^{2}-c^{2})\chi}{b\chi +c} & 0 & 0 \cr
},
\end{equation} 
We set $b-c=1$ to simplify the payoff elements, so we only need to focus on the extortion factor  $\chi$ and the temptation $b$.

\paragraph*{\textbf{Population setting}}

\begin{figure*}[htbp]
  \centering
  \includegraphics[width=0.84\linewidth]{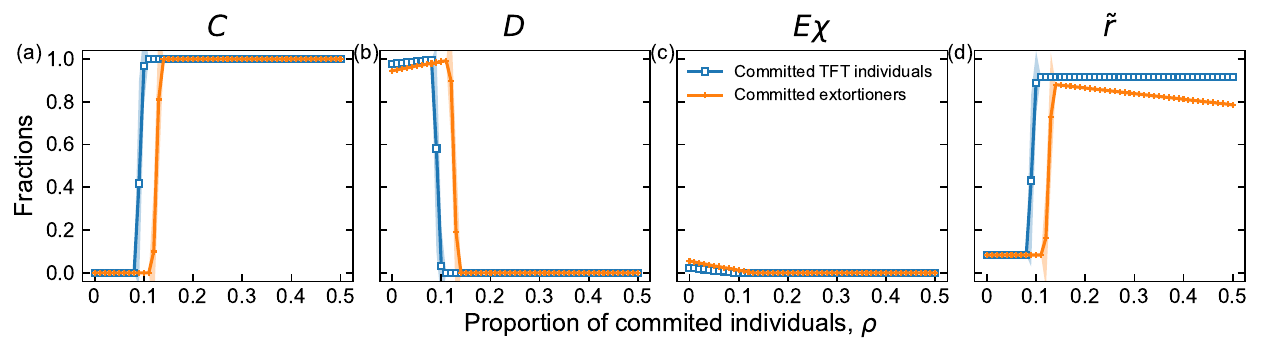}
  \caption{
    \textbf{Committed extortioners exert a critical mass effect on the emergence of cooperation in the Iterated Prisoner's Dilemma game within a well-mixed population.} The figure illustrates the average fraction of (a) \textit{cooperation} ($C$), (b) \textit{defection} ($D$), (c) \textit{extortion} ($E\chi$), and (d) normalized average payoff $\tilde{r}$ for normal players, as a function of the proportion of committed individuals ($\rho$). The results are shown for committed Tit-for-Tat (TFT) individuals ($\chi=1$) and extortioners ($\chi=1.5$), with the parameter $b$ set to $1.1$. The shaded area represents the standard deviation of the results.
  }
  \label{fig:well_cross}
\end{figure*}

Considering a population of size $N$, which consists of a proportion of $\rho$ ($0<\rho<0.5$) committed extortioners who always adopt the $E\chi$ strategy and a proportion of $1-\rho$ normal players. To examine the dynamics and outcomes of the donation game, we consider two typical population structures: well-mixed and structured populations. In a well-mixed population, each player interacts with all $N-1$ others, acquires the cumulative payoff from pairwise IPD game. In a structured population, players are situated on a network where each player occupies a site and can only interact with immediate neighbors. We consider  four typical types of networks: lattice, small-world (SW), Erdős-Rényi (ER), and Bárabasi-Albert (BA) scale-free networks\cite{barabasi1999emergence,szabo2007evolutionary}. The average degree of all networks is fixed as $\left \langle K \right \rangle = 4 $. Lattice and SW networks are both regular networks, but the latter has  a higher clustering coefficient than the former. ER and BA networks are heterogeneous networks, with their degree distributions following Poisson and power-law distributions, respectively.

\paragraph*{\textbf{ Strategy updating} }
To obtain results, we utilize an asynchronous Monte Carlo simulation (MCS) approach. Initially, each normal player holds one of three strategies with equal probability. In each MCS time step, each player acquired cumulative payoffs through playing pairwise games with their direct neighbors. Subsequently, normal players update their strategies using a replicator-like dynamics rule\cite{xu2017extortion}. Specifically, an normal focal individual, says $i$, mimics the strategy of a randomly selected neighbor, says $j$, with a probability $w_{i\leftarrow j}$. This probability is determined by the difference between the cumulative payoffs of the two players:
\begin{equation}
\label{eq:w1}
w_{i\leftarrow j}= \begin{cases}
  \frac{r_j-r_i}{ H \cdot \max(K_j,K_i)} & \text{ if } r_j>r_i \\
  \quad\quad\ 0  & \text{ otherwise }
 \end{cases},
\end{equation}
where $r_x$ and $K_x$ are the player's cumulative payoffs and degree of player $x$, respectively, and $H$ is the maximum possible payoff difference in a pairwise game($H=b+c$ since $\chi>1$). 

The MCS is conducted on networks of size $N=3,000$-$10,000$. We average the results from more than 50 simulations, where each simulation is obtained by averaging the last 3,000 time steps over a total of $5\times 10^{4}$ MCS time steps. This ensures that the fraction of each strategy in the population remains stable.

\section{\label{sec:3} Results}

\begin{figure*}[htbp]
  \centering
  \includegraphics[width=0.78\linewidth]{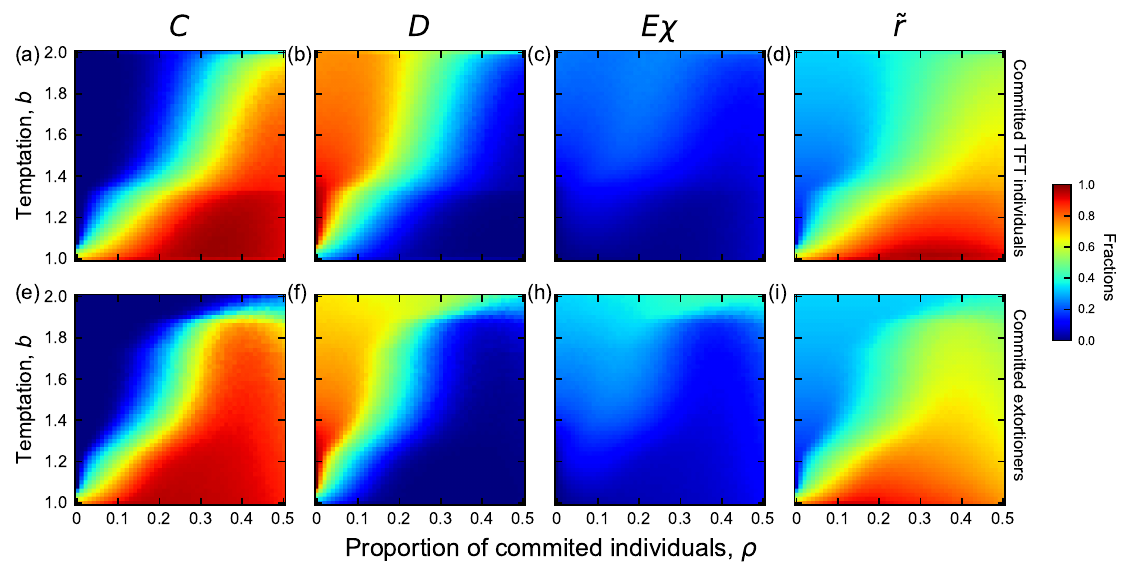}
  \caption{ 
        \textbf{On a regular lattice, a minority of committed extortioners can exert a critical mass effect on the emergence of cooperation in the Iterated Prisoner's Dilemma game under strong imitation scenario, which goes beyond the scope of the one-shot Prisoner's Dilemma game.} The figure illustrates the average fraction (color-coded) of strategies $C$ (cooperate), $D$ (defect), and $E\chi$ (strategy denoted as E, with parameter $\chi$) and the normalized average payoff $\tilde{r}$ for normal players, plotted against the proportion of committed individuals $\rho$ and the temptation parameter $b$. The top panels represent committed Tit-for-Tat (TFT) individuals with $\chi=1$, while the bottom panels depict committed extortioners with $\chi=1.5$.
  }
  \label{fig:grid_phase}
\end{figure*}

\subsection{Results for well-mixed populations}

Before delving into the impact of committed individuals in IPD game, let us revisit the influence of zealots (i.e., committed cooperators) in a one-shot Prisoner's Dilemma game. In the absence of any reciprocal mechanism, defection undeniably emerges as the only Nash equilibrium strategy in such games. Cardillo et al.~\cite{cardillo2020critical} found that introducing zealots into a well-mixed population does not lead to the occurrence of critical mass effects in a one-shot Prisoner's Dilemma game. The research suggests that zealots exhibit a critical mass effect promoting cooperation only under weak selection conditions, wherein cooperative evolution remains independent of payoff differences. These findings suggested that the occurrence of critical mass effects in a one-shot Prisoner's Dilemma game is dependent on the selection intensity parameter. 

In our study, we broaden our scope beyond the zealots and explore whether a group of committed individuals, encompassing both those committed unconditional and conditional cooperators, can demonstrate universal critical mass effect. When taking into account the more common scenario of repeated interactions between individuals, interestingly, we observe the emergence of crucial phenomena triggered by committed individuals in the IPD game, as illustrated in Fig.~\ref{fig:well_cross}. The introduction of a minority of committed conditional cooperators who adopt the TFT or $E\chi$ strategy can have a significant impact on the outcome of cooperation,  In particular, when $\rho \approx 0.09$, the existence of committed TFT individuals ($\rho \approx 0.125$ for committed extortioners) can dramatically shift the outcome from extinction of cooperation ($F_C = 0$) to its dominance ($F_C = 1$). 

Furthermore, we investigated the impact of committed individuals on the payoffs of regular players. The average normalized payoff for normal players is shown in Fig.~\ref{fig:well_cross}(d). When defectors are entirely eliminated, individuals adopting the TFT strategy exhibit behavior equivalent to unconditional cooperators, meaning that committed TFT individuals do not affect the payoffs of normal players. However, the presence of committed extortioners reduces the payoffs for normal cooperators due to the snowdrift relationship between extortioners and cooperators, as described in Eq.\ref{eq:pm}. Therefore, as the proportion of committed extortioners increases, the level of payoffs for normal players decreases as expected.

We further explore whether committed individuals can still induce critical effects under varying levels of selection intensity. In scenarios with strong selection, where players tend to imitate the most successful strategies, the presence of committed extortioners facilitates the evolution of cooperation. The existence of a fraction of committed extortioners in the population diminishes the profit advantage of defectors since they are unable to exploit extortioners as stated in the payoff matrix (Eq.~\ref{eq:pm}). Conversely, in scenarios with weak selection, where players' imitation behavior is independent of payoff differences, committed individuals function more as broadcasters of their own strategies to the entire population. As a result, the critical mass effect triggered by committed extortioners diminishes. However, committed unconditional cooperators can still induce the critical mass effect and achieve pure cooperation under weak selection, as illustrated in Fig.~\ref{fig:well_select}. These findings suggest that committed individuals possess a broad capability to elicit the critical mass effect in repeated games, regardless of the selection intensity. When the extortion factor is low (i.e., $\chi=1.5$), the roles of committed TFT individuals and extortioners are similar. However, as the extortion factors increase, the committed extortioners further reduce the profits of normal cooperators, which weakens the facilitating role of committed extortioners. This attenuation effect eventually leads to the disappearance of critical mass effects, as depicted in Fig.~\ref{fig:well_chi}.

\begin{figure*}[htbp]
  \centering
  \includegraphics[width=0.82\linewidth]{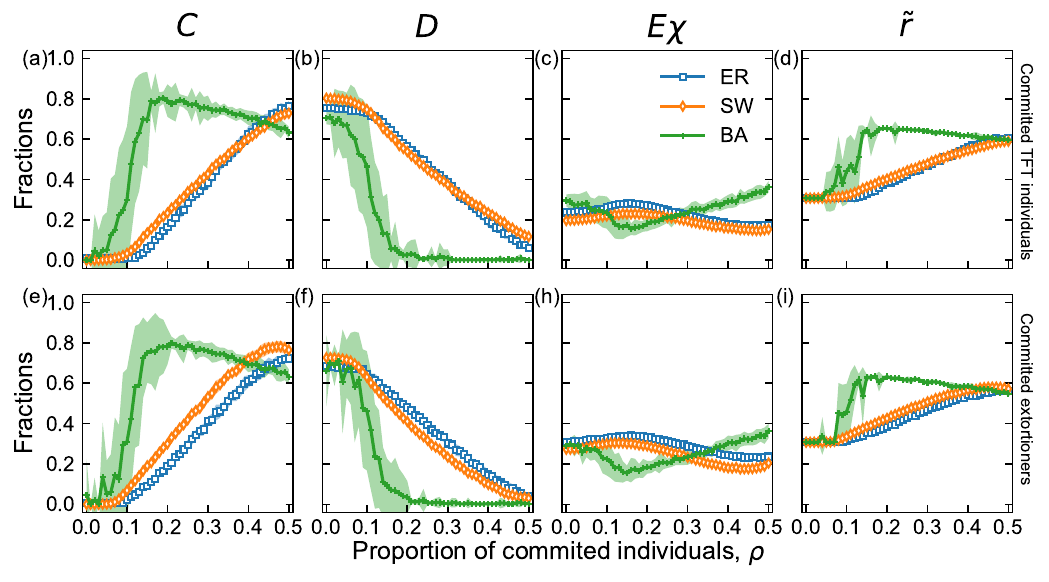}
  \caption{
    \textbf{Committed extortioners can promote collective cooperation on various networks.} Depicted are the average fraction of $C$, $D$, $E\chi$, the normalized average payoff $\tilde{r}$ for normal players as a function of the proportion of committed individuals $\rho$ for committed TFT individuals ($\chi=1$) and extortioners ($\chi=1.5$). The parameters are set to $N=4,000$ and $b=1.8$. We consider three representative network structures: small-world (SW), Erdős-Rényi (ER), and Bárabasi-Albert scale-free (BA) network. The average degree $\left \langle K \right \rangle $ of each network is fixed at 4. When generating the network structure, rewiring probability of the SW network is set to 0.1,the BA network starts with 5 nodes and adds 4 edges per step. The shadowed area represents the standard deviation of the outcome.
  }
  \label{fig:network_cross}
\end{figure*}

\subsection{Results for structured populations}

The occurrence of critical mass effects by zealots in a one-shot Prisoner's Dilemma game, when considering a structured population, depends on specific network structures. This effect is primarily observed in heterogeneous networks, such as scale-free networks Fig.~\cite{cardillo2020critical}. In lattice networks, the role of zealots is counterproductive as it inhibits the evolution of cooperation\cite{matsuzawa2016spatial}. Firstly, we focus on the scenario of a lattice network. Fig.~\ref{fig:grid_phase} suggest that the presence of a small number of committed extortioners can greatly enhance the overall level of cooperation among normal players. In particular, Under high temptation parameters, the presence of even a small fraction of committed extortioners can shift the outcome from complete extinction of cooperation to its existence, similar to observations made in well-mixed populations. Deploying committed TFT individuals also yields similar positive results. Consistent with results from the well-mixed population, Committed extortioners have been found to be effective under strong selection conditions (see Fig.~\ref{fig:grid_select}), whereas committed unconditional cooperators can also trigger critical effects in weak selection scenarios. Furthermore, a high extortion factor negatively impacts their ability to promote cooperation as illustrated in Fig.~\ref{fig:network_chi}. 

It is notable that while a minority of committed individuals can trigger critical mass effects,  introducing more committed individuals may not always efficiently promote cooperation. In particular, under low temptation conditions (i.e., $b \lesssim 1.06$), cooperators are capable of resisting defector invasion even in the absent of committed individuals. Introducing a small number of committed individuals can further enhance cooperation. However, when more committed individuals are introduced (i.e., $\rho \gtrsim 0.38$), regardless of whether they adopt the TFT or the extortion strategy, it can be observed that cooperation declines, see Fig.~\ref{fig:network_chi}(a) and (e). Additionally, the introduction of more committed individuals results in decreased payoff levels for normal players, see Fig.~\ref{fig:network_chi}(d) and (i). This reveals the existence of an optimal $\rho$ that enables the committed individuals to effectively promote cooperation. In subsequent research, we will delve deeper into investigating the underlying reasons behind this phenomenon. 

\begin{figure*}[!t]
  \centering
  \includegraphics[width=0.78\linewidth]{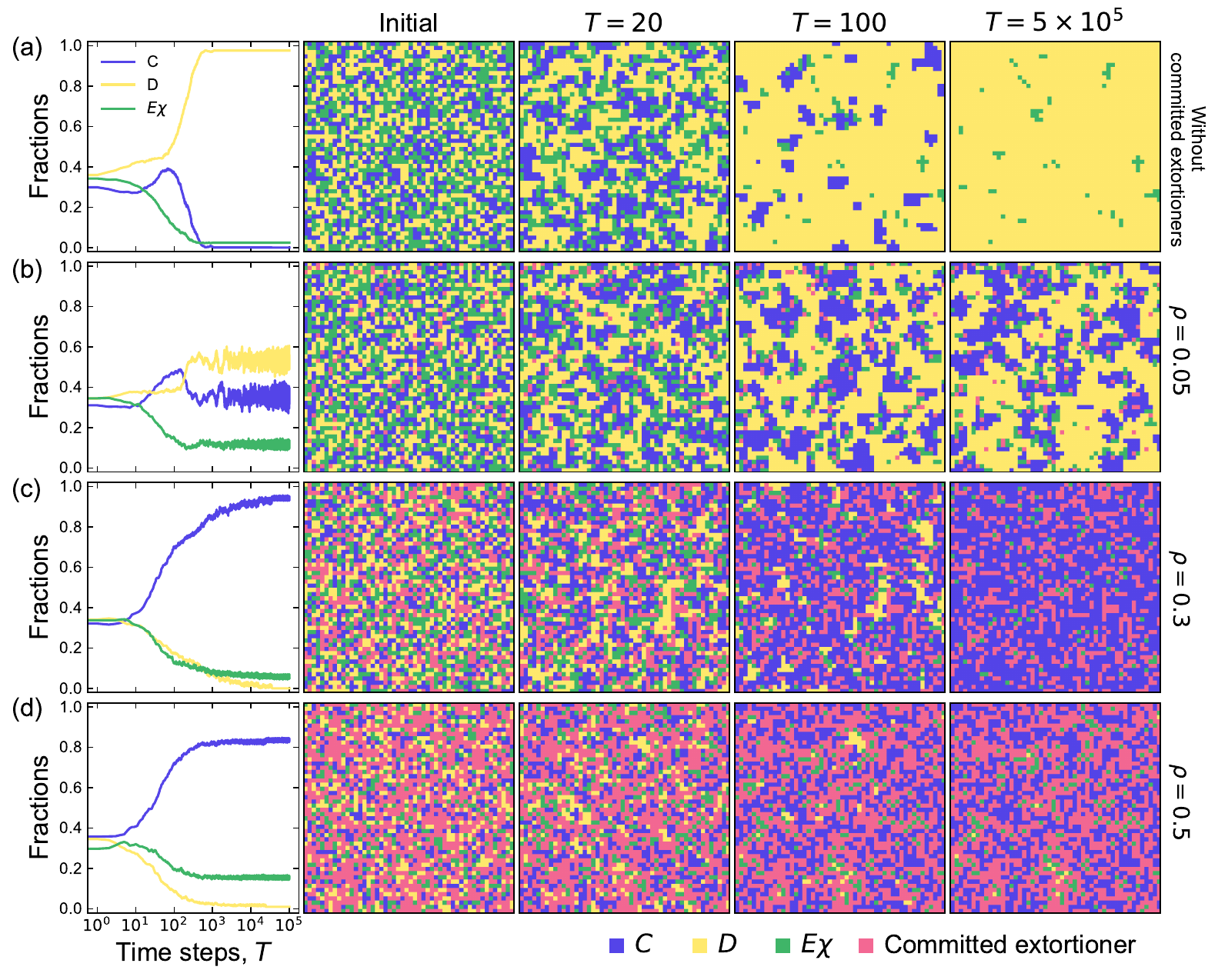}
  \caption{
    \textbf{Too many committed extortioners inhabit cooperation.} Depicted are the fraction of three strategies payoff as a function of time step (left columns), and typical evolutionary snapshots. Blue, yellow, green, and pink marker in the snapshots indicate the cooperator, defector, extortioner, and committed extortioner, respectively. The other parameters are set to $b=1.2$ and $\chi=1.5$.
  }
  \label{fig:snap}
\end{figure*}

To further examine the generality of the critical effect triggered by committed individuals in structured populations, we have considered small-world network, which is a type of homogeneous network exhibiting higher clustering coefficients than lattice network, as well as heterogeneous networks including Erdős-Rényi (ER) and Bárabasi-Albert scale-free (BA) network\cite{barabasi1999emergence, szabo2007evolutionary, dorogovtsev2008critical}. In Fig.~\ref{fig:network_cross}, we report the results obtained from simulations conducted on these networks. Results suggest that the outcomes obtained on SW and ER networks are similar to those observed on lattice network. Specifically, the presence of committed extortioners (or TFT individuals) effectively trigger the critical mass effect and promotes cooperation under strong selection (see Figure \ref{fig:network_cross}).  The critical mass effect of committed unconditional cooperators can also be observed under weak selection(see Fig.~\ref{fig:grid_select} to \ref{fig:er_select}). Remarkably, the fraction of cooperation increases significantly when $\rho=0.06$ in BA network, it is implies that committed extortioners exhibit a critical mass effect on promoting cooperation, and the same is true for committed TFT individuals. We also explore scenarios involving varying levels of selection intensity (see Fig.~\ref{fig:sf_select}), where observe that committed extortioners are effective exclusively under conditions of strong selection. In contrast, committed unconditional cooperators are capable of promoting cooperation regardless of the intensity of selection within the SF network.

In order to validate the robustness of our findings, we examine a scenario where individual strategy imitation follows the pairwise Fermi rule\cite{szabo1998evolutionary}. This rule introduces a level of noise into the strategy update process, reflecting the decision-making dynamics of individuals with bounded rationality. it would be clearer to state "This rule provides a more realistic representation of real-world outcomes compared to the assumption of absolute rationality described in Eq.~\ref{eq:w1}. Significantly, the model consistently produces similar results, even when the strategy update rule is altered, as shown in Fig.~\ref{fig:well_select_fermi} to \ref{fig:sf_select_fermi}. These results demonstrate the universality of critical mass effects, irrespective of the particular strategy update rules employed.

We note that an excessive number of committed extortioners can lead to a lower fraction of $C$ in Fig.~\ref{fig:grid_phase} and \ref{fig:network_cross}. In order to explore the underlying mechanisms governing the impact of committed extortioners on cooperation, we analyze the temporal evolution and representative evolutionary snapshots across various proportions of committed extortioners on a lattice network. We focus on the lattice network as it provides more intuitive insights. During the initial stages of evolution, certain cooperators form clusters and manage to survive within the population, as depicted in the three columns of Fig.~\ref{fig:snap}. In the absence of committed individuals, the formation of cooperation-extortion alliances becomes unfeasible due to a low extortion factor (i.e., $\chi=1.5$) \cite{xu2017extortion}.  Consequently, the fraction of $C$ experiences a significant decline, ultimately leading to the extinction of cooperation, as illustrated in Fig.~\ref{fig:snap}(a). Conversely, even a small proportion of committed extortioners (i.e., $\rho=0.05$) can foster cooperative clusters and uphold cooperation via cooperation-extortion alliances, as shown in Figure X (referring to the specific figure number) and illustrated in Fig.~\ref{fig:snap}(b). When a moderate proportion (i.e., $\rho=0.3$) of committed extortioners is present, the chances likelihood of normal cooperators encountering them increase, thereby promoting the extensive formation of cooperation-extortion alliances, which contributes to the expansion of the cooperative cluster, as indicated in the right panel of Fig.~\ref{fig:snap} (c). As a result, the fraction of $C$ increases, leading to the eventual elimination of defectors. However, when the number of committed extortionists becomes excessive, some non-cooperative individuals become isolated from the cooperators, as depicted in the fourth column of Fig.~\ref{fig:snap} (d). This hinders further expansion of the cooperative cluster, ultimately resulting in an overall lower fraction of cooperation than in the case of $\rho=0.3$.

\section{\label{sec:4}Conclusions}

In conclusion, our study expands the theory of the critical mass of cooperation from the one-shot Prisoner's Dilemma game to the Iterated Prisoner's Dilemma game. Unlike the findings in the one-shot Prisoner's Dilemma game, where the critical mass of cooperation relies on weak imitation strength or scale-free networks under strong imitation strength~\cite{cardillo2020critical}, our extensive Monte Carlo simulations provide compelling evidence for the significant role of committed individuals in driving widespread cooperation in the context of the iterated prisoner's dilemma game. We consistently observed that a minority of committed unconditional cooperators play a pivotal role in generating critical mass effects of cooperation, particularly when the imitation strength is weak, which is in line with previous research~\cite{cardillo2020critical,sharma2023small,guo2023facilitating}. Our results indicate that if the imitation strength is strong, meaning individuals highly value the payoff differences when updating their strategy, a minority of committed conditional cooperators, such as committed extortioners or committed TFT players, can effectively trigger critical mass effects of cooperation.

Importantly, we rigorously tested and validated the robustness of these conclusions across various network topologies and alternative strategy update rules. Regardless of the specific network structures or updating rules considered, our findings consistently highlight the influential effects produced by committed individuals. Therefore, we propose that the critical mass effect of cooperation may represent a universal principle within social systems comprising self-interested individuals.

Furthermore, our findings demonstrate that a small group of committed individuals can significantly facilitate widespread cooperation within a structured population. However, an excessive number of committed individuals can isolate certain cooperators and non-cooperative players, hindering the diffusion of cooperation and diminishing the effectiveness of committed individuals in promoting cooperation as well as the level of payoffs for normal players. Thus, an optimal proportion of committed individuals exists, which maximizes the facilitation of cooperation within a network.

Although the one-shot prisoner's dilemma game, which lacks reciprocity mechanisms, presents the most challenging conditions for cooperation, individuals in this scenario face a simple decision: to cooperate or defect. The limited strategy space confines individuals to two types: committed cooperators or committed defectors. Previous research has shown that committed defectors undermine cooperation, while committed cooperators have limited influence in promoting cooperation. However, by relaxing the assumption of a one-shot game and considering repeated interactions, we can accommodate the diversity of committed individuals' strategies based on memory length~\cite{fudenberg2012slow}. This simple extension allows us to observe the universal principle of a critical mass effect of cooperation under the strong imitation scenario, which extends beyond the one-shot prisoner's dilemma game where defection remains the Nash equilibrium under strong imitation.

Although we have utilized two specific update rules to examine the resilience of the critical mass of cooperation against participants' decision methods, these rules are merely examples based on the disparity in players' payoffs. The key distinction between these rules lies in their nature: the imitation max rule operates as a deterministic update rule, where players can only imitate the strategies of the best-performing players, whereas the Fermi update rule functions as a stochastic rule, allowing for decision errors as players can also imitate strategies of those who perform worse than them. These two rules serve as representatives of stochastic and deterministic processes, respectively. Our findings demonstrate the robustness of the critical mass of cooperation against both stochastic and deterministic rules. Considering the shared underlying principle of these rules, we posit that our results hold true for other update rules as long as they incorporate payoff differences. This is because committed conditional cooperators can trigger a critical mass of cooperation under a strong imitation scenario.

Although one-shot games with some prior information available are more common in realistic scenarios than repeated games or one-shot prisoner's dilemma games without any information, we chose to disregard this situation in our model. Our primary focus was to answer how and why zealots can elicit a universal critical mass effect of cooperation, independent of imitation strength, network topologies, or other updating rules. However, we acknowledge that future studies could explore the critical mass of cooperation within such a framework, especially in the context of indirect reciprocity. By expanding our understanding of the critical mass of cooperation within the framework of indirect reciprocity and reputation dynamics, we can gain a more comprehensive understanding of the factors that influence cooperation in real-world social systems. This could shed light on the mechanisms that facilitate or hinder the formation of critical mass effects of cooperation in complex social systems.

Overall, our model uncovers extensive critical effects initiated by committed individuals and underscores the pivotal role played by a minority of committed individuals in promoting large-scale cooperation. Our results thus enrich the study of the critical mass of cooperation within social systems, especially in systems that contain conflicts between personal interest and collective interest.

\section*{ article information}

\paragraph*{Acknowledgments.} 
We acknowledge the support provided by Major Program of National Fund of Philosophy and Social Science of China (grants no.~22\&ZD158 and 22VRCO49) to L.S.; the grant-in-Aid for Scientific Research from JSPS, Japan, KAKENHI (grant No. JP 20H02314) awarded to J.T.; a JSPS Postdoctoral Fellowship Program for Foreign Researchers (grant no. P21374), an accompanying Grant-in-Aid for Scientific Research from JSPS KAKENHI (grant no. JP 22F31374), and the NNSFC grant no. 11931015) to C.S. as a co-investigator; Yunnan Provincial Department of Education Science Research Fund Project (project no. 2023Y0619) to Z.H.

\paragraph*{Author contributions.} 
Z.H. and C.S. conceptualized, designed the study; Z.H. and C.S. performed simulations and wrote the initial draft of the manuscript; L.S. and J.T. provided overall project supervision; All authors read and approved the final manuscript.

\paragraph*{Conflict of interest.} Authors declare no conflict of interests.





\section*{Appendix A}
\label{sec:sample:appendix}

\setcounter{figure}{0}
\renewcommand{\thefigure}{A\arabic{figure}}
\renewcommand\thesubsection{A\arabic{subsection}}

\setcounter{equation}{0}
\renewcommand\theequation{a\arabic{equation}}

\begin{figure}[ht!]
  \centering
  \includegraphics[width=0.98\linewidth]{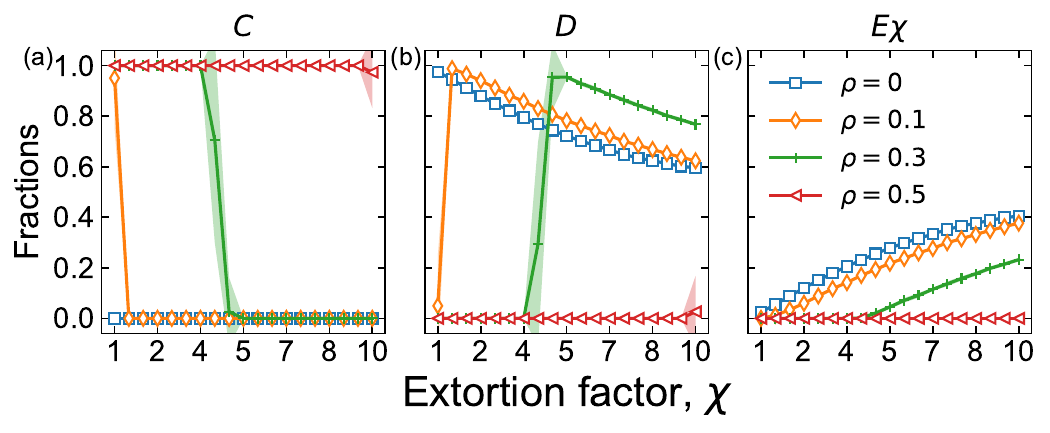}
  \caption{
    \textbf{As the extortion factor increases, the committed extortioner's role in promoting cooperation diminishes in well-mixed population.} Depicted are the average fraction of $C$, $D$, and $E\chi$ among normal players as a function of the extortion factor $\chi$ for committed extortioners. The parameters are set to $b=1.1$. The shadowed area represents the standard deviation of the outcome.
  }
  \label{fig:well_chi}
\end{figure}

\begin{figure}[ht!]
  \centering
\includegraphics[width=0.98\linewidth]{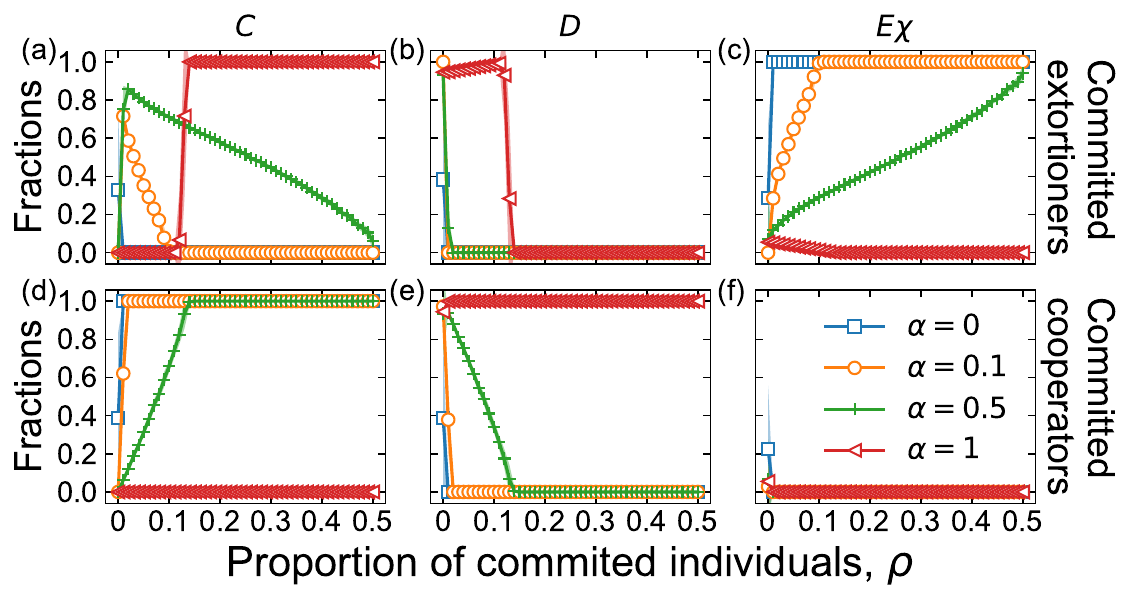}
  \caption{
    \textbf{In well-mixed population, committed extortioners only promote cooperation under strong selection, while committed cooperators only promote cooperation under weak selection.} Depicted are the average fraction of $C$, $D$, and $E\chi$ among normal players as a function of the proportion of committed individuals $\rho$ for committed extortioners individuals and cooperators. Strong selection $\alpha \rightarrow 1$  means that the player's strategy imitation depends on the difference in payoffs, while the individual performs random imitation under weak selection $\alpha \rightarrow 0$. The parameters are set to $b=1.1$ and $\chi=1.5$. The shadowed area represents the standard deviation of the outcome.
  }
  \label{fig:well_select}
 \end{figure}

\subsection{The influence of selection intensity and extortion factor on committed individuals in a mixed population}

Fig.~\ref{fig:well_chi} reports the fractions of three strategies as a function of extortion factor $\chi$ when committed extortioners are introduced into a well-mixed population. While the existence of committed extortioners can weaken the advantage of defectors, an increase in the extortion factor allows extortioners to obtain even greater surplus benefits from their interactions with cooperators, which reduces the payoff of cooperators. As a result, the population becomes more inclined towards extortion strategies. With high extortion factors, the critical mass effect, triggered by a committed extortioners, becomes unsustainable.

To explore the influence of committed individuals under varying selection intensities, we made adjustments to the strategy imitation rule described in Eq.~\ref{eq:w1}. Specifically, the focal player, denoted $i$, mimics the strategy of a randomly selected neighbor, represented by $j$, with a adjusted probability $\tilde{w}_{i\leftarrow j}$:
\begin{equation}
\label{eq:w_selec}
\tilde{w}_{i\leftarrow j} = \frac{1-\alpha}{2} + \alpha \cdot w_{i\leftarrow j},
\end{equation}
where $w_{i\leftarrow j}$ is the imitation probability calculated by replicator-like dynamics rule (Eq.~\ref{eq:w1}), while $\alpha \in[0,1]$ denotes the selection intensity. High selection intensity, where $\alpha = 1$, means that players only mimic strategies that are more successful than their own, whereas low selection intensity, with $\alpha = 0$, results in random imitation.

Figure~\ref{fig:well_select} presents compelling evidence of the critical mass effect exhibited by committed extortioners under strong selection. Even under moderate selection intensity ($\alpha=0.5$), they still demonstrate a critical mass effect, albeit with diminishing impact as $\rho$ increases. Conversely, under weak selection, the presence of committed extortioners facilitates the prevalence of extortion rather than promoting cooperation. In contrast, committed unconditional cooperators significantly enhance cooperation under weak selection and also trigger a critical mass effect.

\subsection{The influence of selection intensity and extortion factor on committed individuals on structured populations}

\begin{figure}[ht!]
  \centering
  \includegraphics[width=0.98\linewidth]{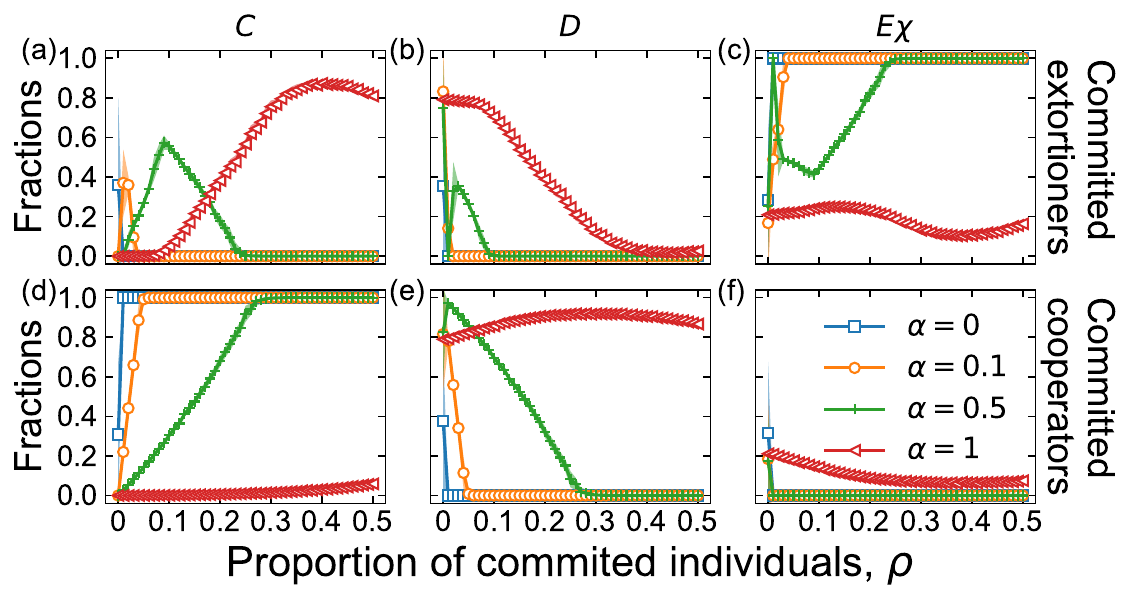}
  \caption{
    \textbf{ The impact of committed individuals on cooperation at various select intensity on lattice network.} Depicted are the average fraction of $C$, $D$, and $E\chi$ among normal players as a function of the proportion of committed individuals $\rho$ for committed extortioners individuals and cooperators. The parameters are set to $b=1.5$ and $\chi=1.5$. The shadowed area represents the standard deviation of the outcome.
  }
  \label{fig:grid_select}
\end{figure}

\begin{figure}[ht!]
  \centering
  \includegraphics[width=0.98\linewidth]{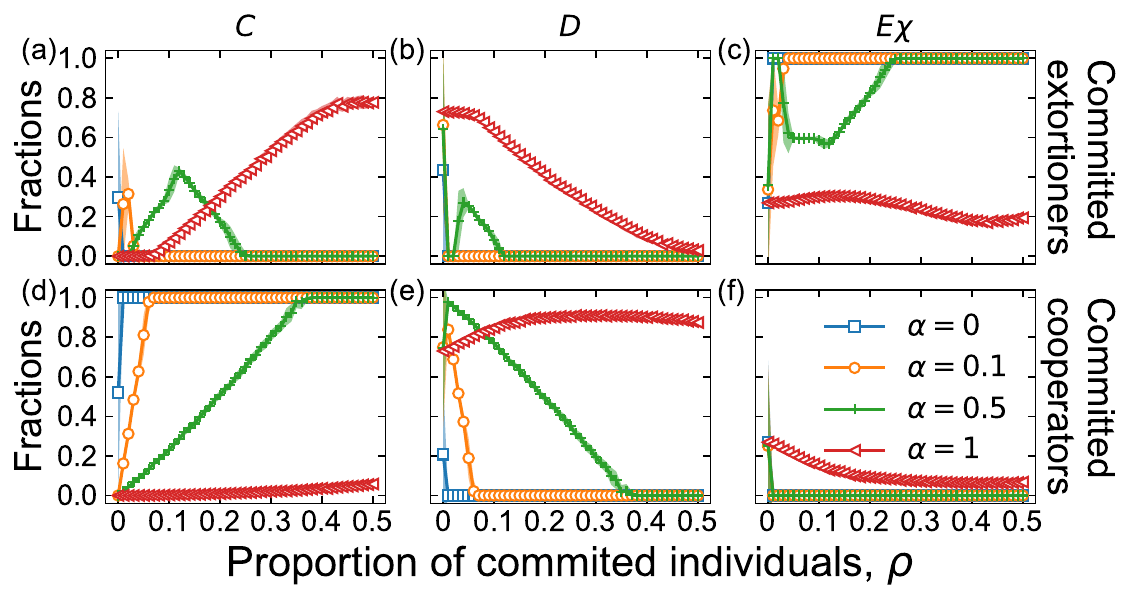}
  \caption{
    \textbf{ The impact of committed individuals on cooperation at various select intensity on small-world network.} Depicted are the average fraction of $C$, $D$, and $E\chi$ among normal players as a function of the proportion of committed individuals $\rho$ for committed extortioners individuals and cooperators. The parameters are set to $b=1.8$, $\chi=1.5$, $N=4,000$ and $\left \langle K \right \rangle = 4 $. The shadowed area represents the standard deviation of the outcome.
  }
  \label{fig:sw_select}
\end{figure}

\begin{figure}[ht!]
  \centering
  \includegraphics[width=0.98\linewidth]{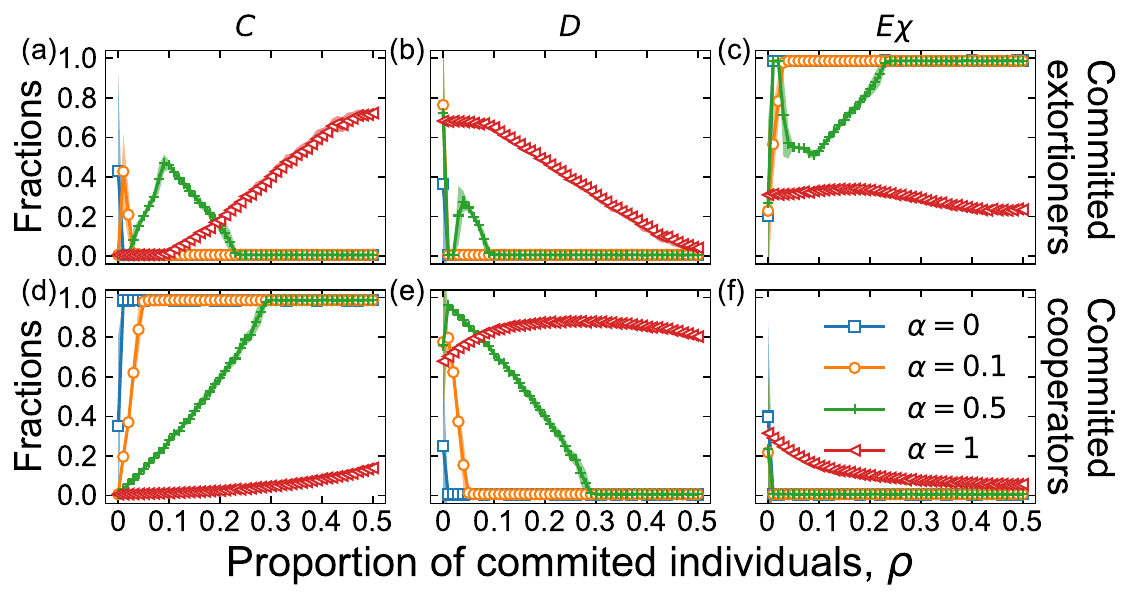}
  \caption{
    \textbf{ The impact of committed individuals on cooperation at various select intensity on Erdős-Rényi network.} Depicted are the average fraction of $C$, $D$, and $E\chi$ among normal players as a function of the proportion of committed individuals $\rho$ for committed extortioners individuals and cooperators. The parameters are set to $b=1.5$, $\chi=1.5$, $N=4,000$ and $\left \langle K \right \rangle = 4 $. The shadowed area represents the standard deviation of the outcome.
  }
  \label{fig:er_select}
\end{figure}

\begin{figure}[ht!]
  \centering
  \includegraphics[width=0.98\linewidth]{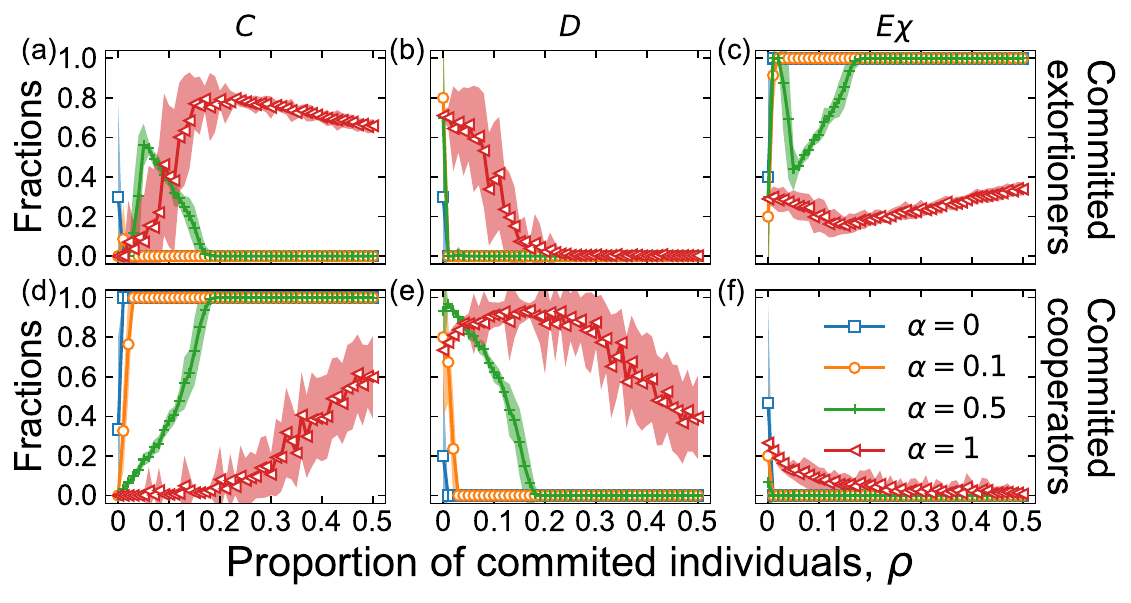}
  \caption{
    \textbf{ The impact of committed individuals on cooperation at various select intensity on Bárabasi-Albert scale-free network.} Depicted are the average fraction of $C$, $D$, and $E\chi$ among normal players as a function of the proportion of committed individuals $\rho$ for committed extortioners individuals and cooperators. The parameters are set to $b=1.8$, $\chi=1.5$, $N=4,000$ and $\left \langle K \right \rangle = 4 $. The shadowed area represents the standard deviation of the outcome.
  }
  \label{fig:sf_select}
\end{figure}

\begin{figure}[ht!]
  \centering
  \includegraphics[width=0.98\linewidth]{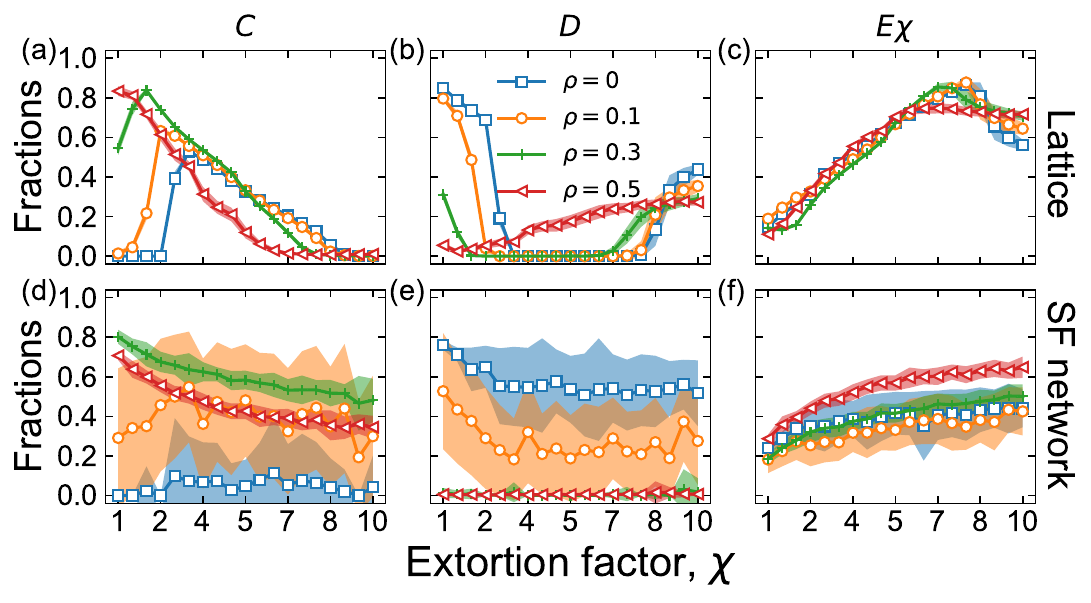}
  \caption{
    \textbf{
      Regardless of the extortion factor, the critical mass effect of committed extortioners persists in scale-free networks.} Depicted are the average fraction of $C$, $D$, and $E\chi$ among normal players as a function of the extortion factor $\chi$ for committed extortioners. In lattice and BA networks, $b$ is set to 1.5 and 1.8, respectively. The shadowed area represents the standard deviation of the outcome.
  }
  \label{fig:network_chi}
 \end{figure}


Based on the findings illustrated in Fig.~\ref{fig:grid_select}-\ref{fig:er_select}, it was observed that the committed individuals had a similar impact on lattice, SW, and ER networks, in terms of select intensity. Presence of committed extortioners can facilitate the evolution of cooperation and exhibit a critical effect effect, which aligns with the results observed in mixed well-behaved populations. This suggests that, under strong selection, committed extortioners can promote cooperation, while committed unconditional cooperators work under weak selection. Furthermore, it was observed that there exists an optimal proportion of committed individuals that effectively facilitates cooperation on these networks.

For the SF network scenario, Fig.~\ref{fig:sf_select} demonstrates that committed extortioners exhibit critical mass effects in promoting cooperation under strong selection. Moreover, we observe that committed unconditional cooperators can also trigger critical mass effect to facilitate the evolution of cooperation among normal players, irrespective of whether the selection is strong or weak. These results reveal that regardless of the strength of selection or the type of network, committed individuals have the potential to extensively trigger critical mass effects.

We also investigated the impact of the extortion factor on committed extortioners within a structured population. Fig.~\ref{fig:network_chi} illustrates that as the extortion factor increases, the ability of committed extortioners to promote cooperation is reduced. However, in a scale-free network, committed extortioners continue to exhibit critical effects regardless of the value of extortion factor.

\subsection{The impact of pair-wise Fermi rule}

\begin{figure}[ht!]
  \centering
  \includegraphics[width=0.98\linewidth]{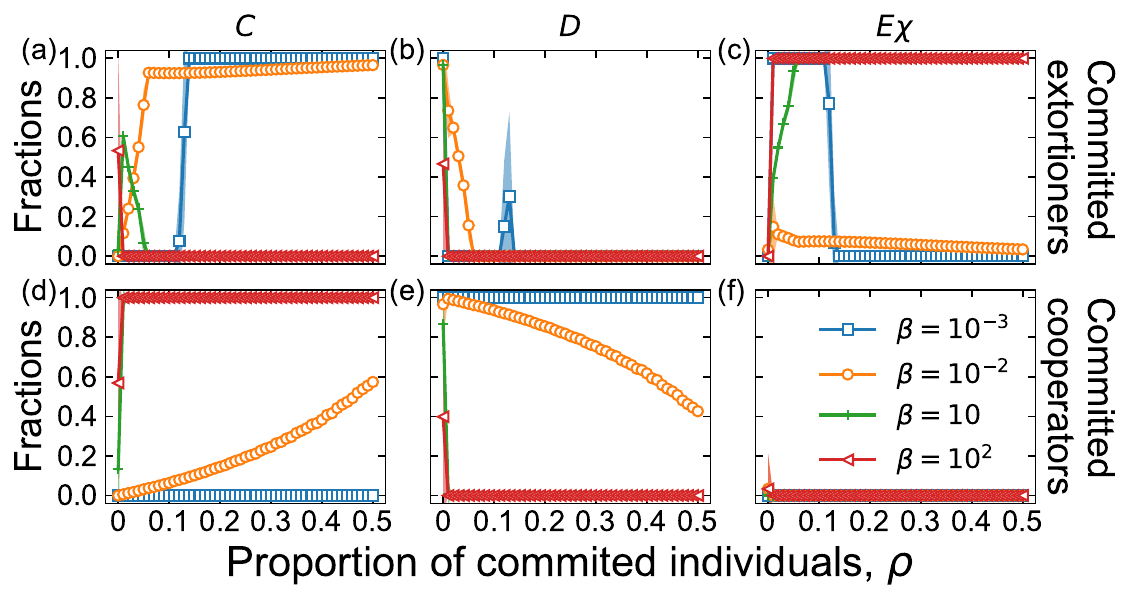}
  \caption{
    \textbf{ The impact of committed individuals on cooperation  using Fermi rule in well-mixed population.} Depicted are the average fraction of $C$, $D$, and $E\chi$ among normal players as a function of the proportion of committed individuals $\rho$ for committed extortioners individuals and cooperators. Strong selection $\beta \rightarrow +\infty$ in Fermi rule  means that the player's strategy imitation depends on the difference in payoffs, while the individual performs random imitation under weak selection $\beta \rightarrow 0$. The parameters are set to $b=1.1$ and $\chi=1.5$. The shadowed area represents the standard deviation of the outcome.
  }
  \label{fig:well_select_fermi}
\end{figure}

\begin{figure}[ht!]
  \centering
  \includegraphics[width=0.9\linewidth]{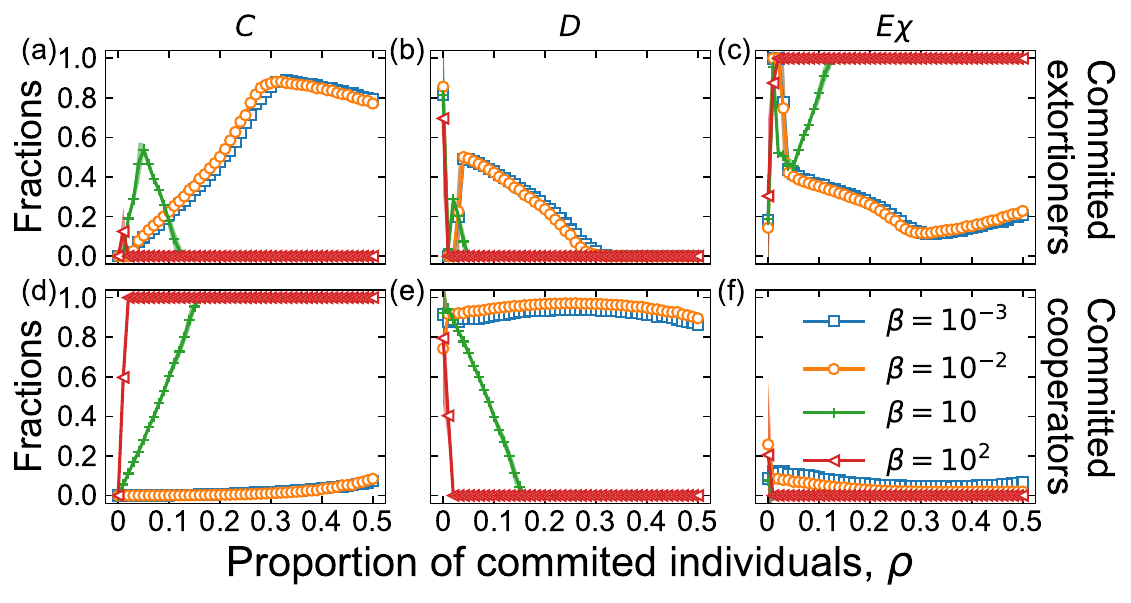}
  \caption{
  \textbf{ The impact of committed individuals on cooperation  using Fermi rule on lattice network.} Depicted are the average fraction of $C$, $D$, and $E\chi$ among normal players as a function of the proportion of committed individuals $\rho$ and temptation $b$ for committed extortioners individuals and cooperators. The parameter is set to $\chi=1.5$. 
  }
  \label{fig:grid_select_fermi}
\end{figure}

\begin{figure}[ht!]
  \centering
  \includegraphics[width=0.98\linewidth]{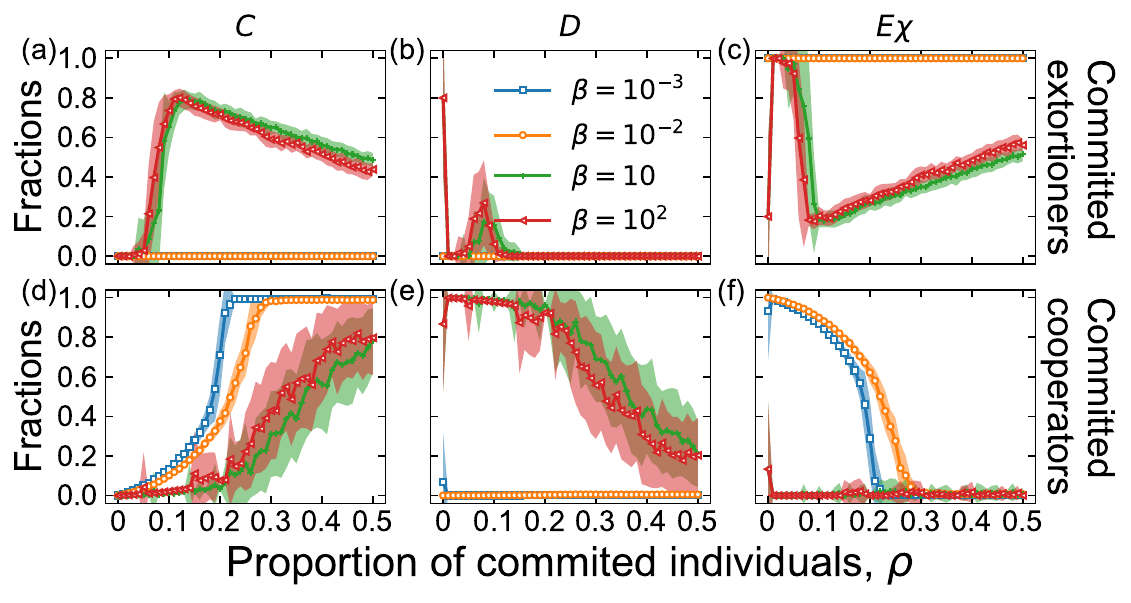}
  \caption{
    \textbf{ The impact of committed individuals on cooperation  using Fermi rule on Bárabasi-Albert scale-free network.} Depicted are the average fraction of $C$, $D$, and $E\chi$ among normal players as a function of the proportion of committed individuals $\rho$ for committed extortioners individuals and cooperators. The parameters are set to $b=1.8$, $\chi=1.5$, $N=3,000$ and $\left \langle K \right \rangle = 4 $. The shadowed area represents the standard deviation of the outcome. The shadowed area represents the standard deviation of the outcome.
  }
  \label{fig:sf_select_fermi}
\end{figure}

\begin{figure*}[ht!]
  \centering
  \includegraphics[width=0.9\linewidth]{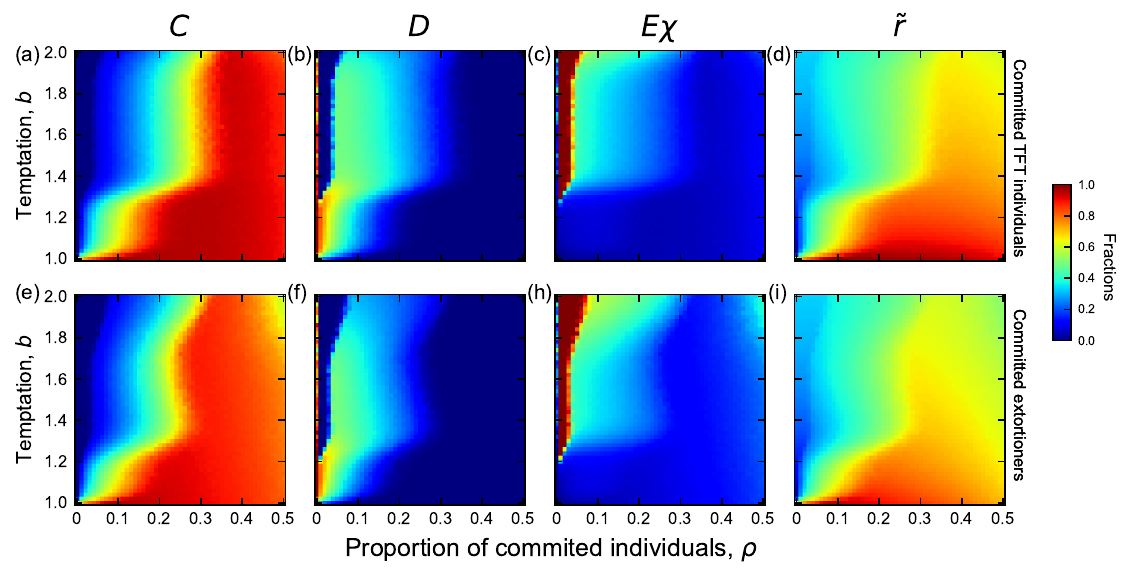}
  \caption{
  \textbf{The impact of committed individuals on cooperation  using Fermi rule on lattice network.} Depicted are the average fraction (color codes) of $C$, $D$, $E\chi$ and the normalized average payoff $\tilde{r}$ for normal players as a function of the proportion of committed individuals $\rho$ and temptation $b$ for committed and TFT individuals ($\chi=1$) and extortioners ($\chi=1.5$). The parameter $\beta$ is set to 0.1.
  }
  \label{fig:grid_b_rho_fermi}
\end{figure*}

Fig.~\ref{fig:well_select_fermi}-\ref{fig:grid_b_rho_fermi} suggest that when player strategy update is governed by the Fermi rule, the role of the committed individuals is robust.

Unlike the scenario described in the main text, in this investigation, the individual's strategy update adheres to replicator-like dynamics rule, where players imitate strategies that are more successful than their own. Specifically, we explore the scenario with the pairwise Fermi rule, which captures the decision-making process of individuals with bounded rationality. In this scenario, a normal focal player $i$ imitates the strategy of a randomly selected neighbor $j$ with a probability of $p_{i\leftarrow j}$:
\begin{equation}
\label{eq:fm}
    p_{i\leftarrow j} =  \frac{1}{1 + \exp\{ (r_i-r_j)/\beta  \}},
\end{equation}
where $r_x$ represents the cumulative payoff of player $x$, $\beta$ denotes the selection intensity. $\beta \rightarrow 0$ indicates that player strategy imitation is completely dependent on by differences in payoffs, while $\beta \rightarrow +\infty$, player strategy imitation tends to become random and unrelated to payoffs. Typically, $\beta$ is set to 0.1.

\clearpage
\bibliography{ref}

\end{document}